\documentclass[a4paper,11pt]{article}
\pdfoutput=1
\usepackage[utf8x]{inputenc}
\usepackage[lmargin=2.5cm,rmargin=2.5cm,tmargin=2.5cm,bmargin=3.5cm]{geometry}
\usepackage[font=small,labelfont=bf,width=0.95\textwidth]{caption}
\usepackage{cite}
\usepackage{subfigure}
\usepackage{amsmath}
\usepackage{amssymb}
\usepackage{fixltx2e}             
\usepackage[T1]{fontenc}          
\usepackage{graphicx}             
\usepackage{feynmp}
\usepackage{booktabs}             
\usepackage{multicol}
\usepackage{etoolbox}
\patchcmd{\thebibliography}{\section*{\refname}}
    {\begin{multicols}{2}[\section*{\refname}]}{}{}
\patchcmd{\endthebibliography}{\endlist}{\endlist\end{multicols}}{}{}
\usepackage[pdftitle={Renormalization of vacuum expectation values in
  spontaneously broken gauge theories: Two-loop results},
pdfauthor={Marcus Sperling,Dominik Stockinger,Alexander Voigt},
pdfkeywords={Renormalization,Vacuum,VEV,MSSM,NMSSM,E6SSM}, bookmarks=true, linktocpage]{hyperref}

\allowdisplaybreaks

\DeclareGraphicsExtensions{.pdf}  

\newcommand{\fmfvcenter}[1]{\vcenter{\hbox{\fmfreuse{#1}}}}	

\newcommand{\Lagr}{\mathcal{L}}				

\DeclareMathOperator{\Tr}{Tr}

\DeclareMathAlphabet{\mathitbf}{OML}{cmm}{b}{it}	

\newcommand{\ESSM}{E\texorpdfstring{\textsubscript{6}}{6}SSM}
\newcommand{\MSbar}{\ensuremath{\overline{\text{MS}}}}
\newcommand{\DRbar}{\ensuremath{\overline{\text{DR}}}}

\newcommand{\GenGauge}[3]{T_{#1 #2}^{#3}}		

\newcommand{\YukaGauge}[3]{Y_{#1 #2}^{#3}}			

\newcommand{\MassScaGauge}[2]{m_{#1 #2}^2} 
\newcommand{\TriScaGauge}[3]{h_{#1 #2 #3}}
\newcommand{\QuadScaGauge}[4]{\lambda_{#1 #2 #3 #4}}

\newcommand{\VecGaugeDo}[2]{V_{#2}^{#1}}

\newcommand{\ScaGauge}[1]{\varphi_{#1}}
\newcommand{\ScaBackGauge}[1]{\hat{\varphi}_{#1}}
\newcommand{\BrstBackGauge}[1]{\hat{q}_{#1}}
\newcommand{\VevGauge}[1]{\hat{v}_{#1}}
\newcommand{\GhoGauge}[1]{c^{#1}}
\newcommand{\AntiGhoGauge}[1]{\bar{c}^{#1}}

\newcommand{\SpinorDo}[2]{\psi_{#1 #2}}				
\newcommand{\SpinorUp}[2]{\psi_{#1}^{#2}}			
\newcommand{\AdjSpinorUp}[2]{\bar{\psi}_{#1}^{\dot{#2}}}	
\newcommand{\MassSpinor}[2]{\left(m_f\right)_{#1 #2}}		

\newcommand{\CasiGenRep}[2]{C_{#1}^2(#2)}

\newcommand{\CasiScaGauge}[2]{C_{#1 #2}^{2}(\mathrm{S})}
 
\newcommand{\CasiAdjGauge}[2]{C_{2}^{#1 #2}(\mathrm{G})}
\newcommand{\DykinScaGauge}[2]{S_{2}^{#1 #2}(\mathrm{S})}
\newcommand{\DykinFermiGauge}[2]{S_{2}^{#1 #2}(\mathrm{F})}

\newcommand{\YukaInvGauge}[2]{Y_{#1 #2}^{2}(\mathrm{S})}        

\newcommand{\BetaFkt}[2]{\beta^{(#1)}(#2)}

\title{\bf Renormalization of vacuum expectation values in
  spontaneously broken gauge theories: Two-loop results}
\author{Marcus Sperling, Dominik St\"ockinger, Alexander Voigt\\[2em]
{\sl Institut f\"ur Kern- und Teilchenphysik,
TU Dresden, Dresden, Germany}}
\date{}

\begin{document}


\newsavebox{\feynmanrules}
\sbox{\feynmanrules}{
\begin{fmffile}{Feynman/FeynmanRules} 
  \fmfset{thin}{.8pt}
  \fmfset{wiggly_len}{2mm}
  \fmfset{dash_len}{2.5mm}
  \fmfset{dot_size}{1thick}
  \fmfset{arrow_len}{2.0mm}
  \fmfset{curly_len}{2.5mm}
  \fmfset{dot_len}{1.6pt}

\begin{fmfgraph}(80,60)
  \fmfkeep{ScaSelfCT}
  \fmfleft{v1}
  \fmfright{v2}
  \fmf{dashes}{v1,c}
  \fmf{dashes}{v2,c}
  \fmfct{c}
\end{fmfgraph}

\begin{fmfgraph}(80,60)
  \fmfkeep{ScaSelfBlob}
  \fmfleft{v1}
  \fmfright{v2}
  \fmf{dashes}{v1,c}
  \fmf{dashes}{c,v2}
  \fmfv{decor.shape=circle,decor.filled=shaded,decor.size=20}{c}
\end{fmfgraph}

\begin{fmfgraph*}(80,60)
  \fmfkeep{TwoPtCTQandBRST}
  \fmfleft{v1}
  \fmfright{v2}
  \fmf{dbl_dashes,label=$\BrstBackGauge{a}$,label.side=left}{v1,c1}
  \fmf{dashes,label=$K_{\ScaGauge{b}}$,label.side=left}{c2,v2}
  \fmf{phantom,tension=5}{c1,c2}
  \fmfct{c1}  
  \fmfv{decor.shape=square,decor.filled=empty,decor.size=8}{c2}
\end{fmfgraph*}

\begin{fmfgraph*}(80,60)
  \fmfkeep{TwoPt2LCTQandBRST}
  \fmfleft{v1}
  \fmfright{v2}
  \fmf{dbl_dashes,label=$\BrstBackGauge{a}$,label.side=left}{v1,c1}
  \fmf{dashes,label=$K_{\ScaGauge{b}}$,label.side=left}{c2,v2}
  \fmf{phantom,tension=5,label=$\delta\hat{Z}^{(2)}$,label.side=right}{c1,c2}
  \fmfv{decor.shape=square,decor.filled=empty,decor.size=8}{c2}
  \fmfct{c1} 
\end{fmfgraph*}

\begin{fmfgraph*}(100,75)
  \fmfkeep{TwoPtSca1Gh1}
  \fmfleft{v1}
  \fmfright{v2}
  \fmf{dbl_dashes,label=$\BrstBackGauge{a}$,label.side=left}{v1,c1}
  \fmf{dashes,label=$K_{\ScaGauge{b}}$,label.side=left}{c2,v2}
  \fmfv{decor.shape=square,decor.filled=empty,decor.size=8}{c2}
  \fmf{ghost,left,tension=0.3}{c1,c2}
  \fmf{dashes,right,tension=0.3}{c1,c2}
  \fmfdot{c1}
\end{fmfgraph*}

\begin{fmfgraph*}(100,75)
    \fmfkeep{TwoPt2LSca1Gh1Spi2}
    \fmfleft{v1}
    \fmfright{v2}
    \fmf{dbl_dashes,label=$\BrstBackGauge{a}$,label.side=left}{v1,c1}
    \fmf{dashes,label=$K_{\ScaGauge{b}}$,label.side=left}{c2,v2}
    \fmf{ghost,left,tension=0.3}{c1,c2}
    \fmf{dashes,tension=1}{c1,c3}
    \fmf{dashes,tension=1}{c4,c2}
    \fmf{plain,left,tension=0.5}{c3,c4}
    \fmf{plain,right,tension=0.5}{c3,c4}
    \fmfv{decor.shape=square,decor.filled=empty,decor.size=8}{c2}
    \fmfdot{c1,c3,c4}
\end{fmfgraph*}

\begin{fmfgraph*}(80,60)
    \fmfkeep{TwoPtCT2LSca1Gh1Spi2}
    \fmfleft{v1}
    \fmfright{v2}
    \fmf{dbl_dashes,label=$\BrstBackGauge{a}$,label.side=left}{v1,c1}
    \fmf{dashes,label=$K_{\ScaGauge{b}}$,label.side=left}{c2,v2}
    \fmf{ghost,left,tension=0.3}{c1,c2}
    \fmf{dashes,tension=0.4}{c1,c3}
    \fmf{dashes,tension=0.4}{c4,c2}
    \fmfv{decor.shape=square,decor.filled=empty,decor.size=8}{c2}
    \fmfdot{c1}
    \fmfct{c3}
    \fmf{phantom,tension=10,label=$\delta Z^{(1)}$,label.side=right}{c3,c4}
\end{fmfgraph*}

\begin{fmfgraph*}(80,60)
  \fmfkeep{ThreePtBrstSource1Sca1Gh1}
  \fmfleft{v1}
  \fmfright{v2,v3}
  \fmf{dashes,label=$K_{\ScaGauge{a}}$,label.side=left}{v1,c}
  \fmfv{decor.shape=square,decor.filled=empty,decor.size=8}{c}
  \fmf{ghost,label=$\GhoGauge{A}$,label.side=right}{v3,c}
  \fmf{dashes,label=$\ScaGauge{b}$,label.side=left}{c,v2}
\end{fmfgraph*}

\begin{fmfgraph*}(80,60)
  \fmfkeep{ThreePtBrstBack1Sca1AntiGh1}
  \fmfleft{v1}
  \fmfright{v2,v3}
  \fmf{dbl_dashes,label=$\BrstBackGauge{a}$,label.side=left}{v1,c}
  \fmf{ghost,label=$\AntiGhoGauge{A}$,label.side=left}{c,v3}
  \fmf{dashes,label=$\ScaGauge{b}$,label.side=left}{c,v2}
  \fmfdot{c}
\end{fmfgraph*}

\begin{fmfgraph*}(80,60)
  \fmfkeep{TwoPtBrstBackTwoLoopScaSelf}
  \fmfleft{i}
  \fmfright{o}
  \fmf{dbl_dashes,label=$\BrstBackGauge{a}$,label.side=left}{i,v1}
  \fmf{dashes,label=$K_{\ScaGauge{b}}$,label.side=left}{v2,o}
  \fmfdot{v1}
  \fmfv{decor.shape=square,decor.filled=empty,decor.size=8}{v2}
  \fmf{dashes,tension=0.3}{v1,v3}
  \fmf{dashes,tension=0.3}{v3,v2}
  \fmfv{decor.shape=circle,decor.filled=shaded,decor.size=15}{v3}
  \fmf{ghost,left,tension=0.3}{v1,v2}
\end{fmfgraph*}

\begin{fmfgraph*}(80,60)
  \fmfkeep{TwoPtBrstBackTwoLoopScaSelfCT}
  \fmfleft{i}
  \fmfright{o}
  \fmf{dbl_dashes,label=$\BrstBackGauge{a}$,label.side=left}{i,v1}
  \fmf{dashes,label=$K_{\ScaGauge{b}}$,label.side=left}{v2,o}
  \fmfdot{v1}
  \fmfv{decor.shape=square,decor.filled=empty,decor.size=8}{v2}
  \fmf{dashes,tension=0.3}{v1,v3}
  \fmf{dashes,tension=0.3}{v3,v2}
  \fmfct{v3}
  \fmf{ghost,left,tension=0.3}{v1,v2}
\end{fmfgraph*}

\begin{fmfgraph*}(80,60)
  \fmfkeep{TwoPtBrstBackTwoLoopGhoSelf}
  \fmfleft{i}
  \fmfright{o}
  \fmf{dbl_dashes,label=$\BrstBackGauge{a}$,label.side=left}{i,v1}
  \fmf{dashes,label=$K_{\ScaGauge{b}}$,label.side=left}{v2,o}
  \fmfdot{v1}
  \fmfv{decor.shape=square,decor.filled=empty,decor.size=8}{v2}
  \fmf{ghost,tension=0.1}{v1,v3}
  \fmf{ghost,tension=0.1}{v3,v2}
  \fmfv{decor.shape=circle,decor.filled=shaded,decor.size=15}{v3}
  \fmf{dashes,left,tension=0.3}{v1,v2}
\end{fmfgraph*}

\begin{fmfgraph*}(80,60)
  \fmfkeep{TwoPtBrstBackTwoLoopGhoSelfCT}
  \fmfleft{i}
  \fmfright{o}
  \fmf{dbl_dashes,label=$\BrstBackGauge{a}$,label.side=left}{i,v1}
  \fmf{dashes,label=$K_{\ScaGauge{b}}$,label.side=left}{v2,o}
  \fmfdot{v1}
  \fmfv{decor.shape=square,decor.filled=empty,decor.size=8}{v2}
  \fmf{ghost,tension=0.1}{v1,v3}
  \fmf{ghost,tension=0.1}{v3,v2}
  \fmfct{v3}
  \fmf{dashes,left,tension=0.3}{v1,v2}
\end{fmfgraph*}

\begin{fmfgraph*}(80,60)
  \fmfkeep{TwoPtBrstBackTwoLoopVecExchange}
  \fmfleft{i}
  \fmfright{o}
  \fmftop{t}
  \fmfbottom{b}
  \fmf{phantom,tension=1}{t,v3}
  \fmf{phantom,tension=1}{b,v4}
  \fmf{dbl_dashes,label=$\BrstBackGauge{a}$,label.side=left}{i,v1}
  \fmf{dashes,label=$K_{\ScaGauge{b}}$,label.side=left}{v2,o}
  \fmfdot{v1}
  \fmfv{decor.shape=square,decor.filled=empty,decor.size=8}{v2}
  \fmf{ghost,tension=0.3}{v1,v3}
  \fmf{ghost,tension=0.3}{v3,v2}
  \fmf{dashes,tension=0.3}{v1,v4}
  \fmf{dashes,tension=0.3}{v4,v2}
  \fmf{photon,tension=0}{v3,v4}
  \fmfdot{v1,v3,v4}
\end{fmfgraph*}

\begin{fmfgraph*}(80,60)
  \fmfkeep{TwoPtBrstBackTwoLoopScaExchange}
  \fmfleft{i}
  \fmfright{o}
  \fmftop{t}
  \fmfbottom{b}
  \fmf{phantom,tension=1}{t,v3}
  \fmf{phantom,tension=1}{b,v4}
  \fmf{dbl_dashes,label=$\BrstBackGauge{a}$,label.side=left}{i,v1}
  \fmf{dashes,label=$K_{\ScaGauge{b}}$,label.side=left}{v2,o}
  \fmfdot{v1}
  \fmfv{decor.shape=square,decor.filled=empty,decor.size=8}{v2}
  \fmf{ghost,tension=0.3}{v1,v3}
  \fmf{ghost,tension=0.3}{v3,v2}
  \fmf{dashes,tension=0.3}{v1,v4}
  \fmf{dashes,tension=0.3}{v4,v2}
  \fmf{dashes,tension=0}{v3,v4}
  \fmfdot{v1,v3,v4}
\end{fmfgraph*}

\begin{fmfgraph*}(80,60)
  \fmfkeep{TwoPtBrstBackTwoLoopGhoExchange}
  \fmfleft{i}
  \fmfright{o}
  \fmftop{t}
  \fmfbottom{b}
  \fmf{phantom,tension=1}{t,v3}
  \fmf{phantom,tension=1}{b,v4}
  \fmf{dbl_dashes,label=$\BrstBackGauge{a}$,label.side=left}{i,v1}
  \fmf{dashes,label=$K_{\ScaGauge{b}}$,label.side=left}{v2,o}
  \fmfdot{v1}
  \fmfv{decor.shape=square,decor.filled=empty,decor.size=8}{v2}
  \fmf{ghost,tension=0.3}{v1,v3}
  \fmf{ghost,tension=0.3}{v4,v2}
  \fmf{dashes,tension=0.3}{v1,v4}
  \fmf{dashes,tension=0.3}{v3,v2}
  \fmf{ghost,tension=0}{v3,v4}
  \fmfdot{v1,v3,v4}
\end{fmfgraph*}

\begin{fmfgraph*}(80,60)
  \fmfkeep{TwoPtBrstBackTwoLoopCTqVertex}
  \fmfleft{i}
  \fmfright{o}
  \fmf{dbl_dashes,label=$\BrstBackGauge{a}$,label.side=left}{i,v1}
  \fmf{dashes,label=$K_{\ScaGauge{b}}$,label.side=left}{v2,o}
  \fmfct{v1}
  \fmfv{decor.shape=square,decor.filled=empty,decor.size=8}{v2}
  \fmf{ghost,left,tension=0.3}{v1,v2}
  \fmf{dashes,right,tension=0.3}{v1,v2}
\end{fmfgraph*}

\begin{fmfgraph*}(80,60)
  \fmfkeep{TwoPtBrstBackTwoLoopCTbrstVertex}
  \fmfleft{i}
  \fmfright{o}
  \fmf{dbl_dashes,label=$\BrstBackGauge{a}$,label.side=left}{i,v1}
  \fmf{dashes,label=$K_{\ScaGauge{b}}$,label.side=left}{v2,o}
  \fmfdot{v1}
  \fmfv{decor.shape=square,decor.filled=empty,decor.size=8}{v2}
  \fmf{phantom,tension=5}{v2,v3}
  \fmfct{v3}
  \fmf{ghost,left,tension=0.3}{v1,v3}
  \fmf{dashes,right,tension=0.3}{v1,v3}
\end{fmfgraph*}

\begin{fmfgraph*}(80,60)
  \fmfkeep{TwoPtBrstBackTwoLoopScaSelf1Vec1Sca}
  \fmfleft{i}
  \fmfright{o}
  \fmf{dbl_dashes,label=$\BrstBackGauge{a}$,label.side=left}{i,v1}
  \fmf{dashes,label=$K_{\ScaGauge{b}}$,label.side=left}{v2,o}
  \fmfdot{v1,v3,v4}
  \fmfv{decor.shape=square,decor.filled=empty,decor.size=8}{v2}
  \fmf{dashes,tension=0.3}{v1,v3}
  \fmf{dashes,tension=0.15,right}{v3,v4}
  \fmf{photon,tension=0.15,left}{v3,v4}
  \fmf{dashes,tension=0.3}{v4,v2}
  \fmf{ghost,left,tension=0.3}{v1,v2}
\end{fmfgraph*}

\begin{fmfgraph*}(80,60)
  \fmfkeep{TwoPtBrstBackTwoLoopScaSelf2Spinor}
  \fmfleft{i}
  \fmfright{o}
  \fmf{dbl_dashes,label=$\BrstBackGauge{a}$,label.side=left}{i,v1}
  \fmf{dashes,label=$K_{\ScaGauge{b}}$,label.side=left}{v2,o}
  \fmfdot{v1,v3,v4}
  \fmfv{decor.shape=square,decor.filled=empty,decor.size=8}{v2}
  \fmf{dashes,tension=0.3}{v1,v3}
  \fmf{plain,tension=0.15,right}{v3,v4}
  \fmf{plain,tension=0.15,left}{v3,v4}
  \fmf{dashes,tension=0.3}{v4,v2}
  \fmf{ghost,left,tension=0.3}{v1,v2}
\end{fmfgraph*}

\begin{fmfgraph*}(80,60)
  \fmfkeep{TwoPtBrstBackTwoLoopGhoSelf1Vec1Gho}
  \fmfleft{i}
  \fmfright{o}
  \fmf{dbl_dashes,label=$\BrstBackGauge{a}$,label.side=left}{i,v1}
  \fmf{dashes,label=$K_{\ScaGauge{b}}$,label.side=left}{v2,o}
  \fmfdot{v1,v3,v4}
  \fmfv{decor.shape=square,decor.filled=empty,decor.size=8}{v2}
  \fmf{ghost,tension=0.3}{v1,v3}
  \fmf{ghost,tension=0.15,right}{v3,v4}
  \fmf{photon,tension=0.15,left}{v3,v4}
  \fmf{ghost,tension=0.3}{v4,v2}
  \fmf{dashes,left,tension=0.3}{v1,v2}
\end{fmfgraph*}

\end{fmffile}
}


\maketitle

\begin{abstract}
 \noindent %
We complete the two-loop calculation of $\beta$-functions
 for vacuum expectation values (VEVs) in gauge theories by the missing
 $\mathcal{O}(g^4)$-terms.  The full two-loop results are presented
 for generic and supersymmetric theories up to two-loop level in
 arbitrary $R_\xi$-gauge. The results are obtained by means of a
 scalar background field, identical to our previous analysis. As a
 by-product, the two-loop scalar anomalous dimension for generic
 supersymmetric theories is presented. As an application we compute
 the $\beta$-functions for VEVs and $\tan \beta$ in the MSSM, NMSSM,
 and \ESSM.
\end{abstract}
\tableofcontents
\clearpage
\section{Introduction}

The renormalization of vacuum expectation values (VEVs) in general
gauge theories with $R_\xi$-gauge has been studied in our earlier
work~\cite{Sperling:2013eva}.  We showed that in $R_\xi$-gauge the
VEVs renormalize differently from the respective scalar fields and
explained the origin and behaviour of this difference.  We computed
VEV-counterterms and $\beta$-functions at one-loop and leading
two-loop level.  The purpose of this subsequent paper is to
complete the two-loop renormalization of VEVs in general gauge
theories and generic supersymmetric theories.

The renormalization of a VEV $v$ can generically be written in the two
equivalent forms
  \begin{align}
  v\to v+\delta v &= \sqrt{Z}\left( v +\delta\bar{v}\right),
  \label{DeltaVDecomposition}
  \end{align}
with $\sqrt{Z}$ being the field renormalization constant of the corresponding scalar
field. The main insight of Ref.~\cite{Sperling:2013eva} has been that
$\delta\bar{v}$ can be interpreted by the field renormalization $ \sqrt{\hat{Z}}
$ of a suitable chosen scalar background field.  Thus, a simple computation
becomes possible in terms of a single two-point function.

In the present paper we address the following points:
\begin{enumerate}
\item The missing two-loop terms of the order $g^4$ in
  $\sqrt{\hat{Z}}$ are computed and the complete two-loop VEV
  $\beta$-function for 
  general gauge theories with $R_\xi$ gauge fixing can be provided.
\item Gauge kinetic mixing in case of several $U(1)$
  gauge factors is taken into account in the computation of the
  $g^4$ terms.
\item The complete results are specialised to general supersymmetric
  theories in the \DRbar\ scheme.
\item As a by-product the anomalous dimension $\gamma^{(2)}$
  for generic $N=1$ supersymmetric theories is derived in \DRbar\ for arbitrary
  values of $\xi$.
\item As application, the concrete results for anomalous dimensions and
$\beta$-functions of VEVs and $\tan\beta$ are provided in the well-known
supersymmetric models MSSM, NMSSM, and \ESSM. These results can be readily
applied in practical applications. Moreover, they highlight various
characteristic features of the general results.
\end{enumerate}
This paper is organized as follows: Sec.~\ref{sec:2_theory} provides a
brief summary of the formalism and notation.  Sec.~\ref{sec:3_methods}
is centred on the computation of the full two-loop results for general
gauge theories and supersymmetric theories. The application to the MSSM, NMSSM,
and \ESSM\ is carried out in Sec.~\ref{sec:4_results}.  Generally this
paper provides a complete picture up to two-loop level and summarizes
all relevant expressions, but the one-loop and Yukawa-enhanced
two-loop results have already been published in
\cite{Sperling:2013eva}.

\section{General Gauge Theory and Scalar Background Fields}
\label{sec:2_theory}
The renormalization of vacuum expectations can be cast in an elegant
scheme by employing a scalar background field.  As elaborated in our
previous publication \cite{Sperling:2013eva}, we use the general
setting of real scalar fields $\ScaGauge{a}$, Weyl 2-spinors
$\SpinorDo{p}{\alpha}$, and real (non-abelian) gauge fields
$\VecGaugeDo{A}{\mu}$ in the notation of
\cite{Machacek:1983tz,Machacek:1983fi,Machacek:1984zw,Luo:2002ti}. The
Lagrangian is given as
\begin{align}
  \Lagr_{\mathrm{inv}}= &-\frac{1}{4} F_{\mu \nu}^A F^{A \mu \nu} +
\frac{1}{2} \left(D_\mu \ScaGauge{} \right)_a \left(D^\mu \ScaGauge{} \right)_a
+ i \SpinorUp{p}{\alpha} \sigma_{\alpha \dot{\alpha}}^{\mu} \left(
D_{\mu}^{\dagger} \AdjSpinorUp{}{\alpha}  \right)_p \nonumber \\*
  &- \frac{1}{2!} \MassScaGauge{a}{b} \ScaGauge{a} \ScaGauge{b} - \frac{1}{3!}
\TriScaGauge{a}{b}{c} \ScaGauge{a} \ScaGauge{b} \ScaGauge{c} - \frac{1}{4!}
\QuadScaGauge{a}{b}{c}{d}  \ScaGauge{a} \ScaGauge{b} \ScaGauge{c} \ScaGauge{d} %
  \label{eqn:lagrangian_basic}
\\*
  &- \frac{1}{2} \left[ \MassSpinor{p}{q} \SpinorUp{p}{\alpha}
\SpinorDo{q}{\alpha} + \mathrm{h.c.} \right]  - \frac{1}{2} \left[
\YukaGauge{p}{q}{a} \SpinorUp{p}{\alpha} \SpinorDo{q}{\alpha} \ScaGauge{a} +
\mathrm{h.c.} \right] . \nonumber%
\end{align}%
The VEVs $v_a$ are replaced in this formalism by  scalar background
fields $(\ScaBackGauge{a} + \VevGauge{a})$.  These auxiliary fields
allow to formulate a rigid (global) gauge invariant gauge fixing;
analogous to Ref.~\cite{Kraus:1995jk} the gauge-fixing functional
reads
\begin{align}
 F^A = \partial^{\mu} \VecGaugeDo{A}{\mu}  +i g \xi \xi' \left(\ScaBackGauge{} +
\VevGauge{} \right)_a \GenGauge{a}{b}{A} \ScaGauge{b} \;. %
\label{eqn:Rxi_modified}
\end{align}
By setting $\ScaBackGauge{a}$ to zero, one recovers the gauge theory
in standard $R_\xi$-gauge.  But the inclusion of $\ScaBackGauge{a}$
and the rigid (global) gauge invariant gauge fixing imply that the
following renormalization transformations are sufficient
\begin{subequations}
 \begin{align}
      \ScaGauge{a} &\rightarrow \sqrt{Z}_{a b} \, \ScaGauge{b} \; ,
\label{eqn:RT_scalar} \\
    ( \ScaBackGauge{} + \VevGauge{} )_a &\rightarrow \sqrt{Z}_{a b} \,
\sqrt{\hat{Z}}_{b c} \, ( \ScaBackGauge{} + \VevGauge{} )_c \; .
\label{eqn:RT_background}
 \end{align}
\end{subequations}
An additional VEV counterterm is then prohibited.  In the standard
approach, without background fields, the most generic renormalization
transformation of the scalar fields with shifts reads
\begin{align}
    \ScaGauge{a} + v_{a} \rightarrow \sqrt{Z}_{a b} \left(
      \ScaGauge{b} + v_{b} + \delta \bar{v}_{b} \right) =
    \sqrt{Z}_{ab} \left(\ScaGauge{b} + v_{b} \right) + \delta v_{a}
  \label{eqn_RT_phi_standard}
\end{align}
The two formalisms are equivalent, with the following identifications
\begin{subequations}
\begin{align}
   \delta v_a &=  \left( \sqrt{Z} \sqrt{\hat{Z}} -1 \right)_{ab}
 \VevGauge{b} = \frac{1}{2} \left( \delta Z + \delta\hat{Z} \right)_{ab}
 \VevGauge{b} +  \mathcal{O}(\hbar^2) \; , %
 \label{eqn:VEV_decomposition_1loop}  \\
\delta \bar{v}_a &= \left(\sqrt{\hat{Z}} - 1 \right)_{a b}
 \VevGauge{b} = \frac{1}{2} \delta \hat{Z}_{a b} \VevGauge{b} +
 \mathcal{O}(\hbar²).
\end{align}
\end{subequations}
As a result, the $\beta$ function of the VEV can be obtained as
\begin{align}
  \beta(v_a) &= \left( \gamma_{ab} + \hat{\gamma}_{ab} \right) v_b \;.
\label{eqn:VEV_beta_def}
\end{align}
with the anomalous dimensions $\gamma$ and $\hat{\gamma}$
corresponding to the field renormalizations $\sqrt{Z}$ and
$\sqrt{\hat{Z}}$, respectively.

One of the main results of Ref.~\cite{Sperling:2013eva} was that the
computation of $\delta\hat Z$ can be reduced to the very simple,
unphysical two-point function
\begin{align}
  \Gamma_{\BrstBackGauge{a},K_{\ScaGauge{b}}}^{\text{CT},(n)} = -
  \frac{i}{2} \delta \hat{Z}_{b a}^{(n)} \; .
\end{align}
Here $K_{\ScaGauge{b}}$ are the sources of the BRS transformation of
the scalar field, and $\BrstBackGauge{a}$ is the BRS transformation of
$\ScaBackGauge{a}$.  Both of these unphysical fields appear in a very
simple and well prescribed way in the Lagrangian.

Our formalism is independent of the actual value assigned to $\hat
v_a$.  We can therefore choose $\hat v_a$ as the minimum of the full
loop-corrected scalar potential.  Hence, our $\beta$-functions describe
the running of the full VEV, which is required, for example, in many
supersymmetry applications such as spectrum generators
\cite{Barger:1993gh,Pierce:1996zz}.  Note that this running VEV has
to be distinguished from other definitions used for example in the
Standard Model \cite{Bednyakov:2013cpa,Jegerlehner:2012kn}, which
corresponds to the VEV defined explicitly in terms of the running
tree-level potential parameters
\begin{align}
  v(\mu) = \sqrt{\frac{m^2(\mu)}{\lambda(\mu)}}.
\end{align}
Ref.~\cite{Bednyakov:2013cpa} contains a diagram exposing the difference
in the running between the different definitions.

\section{Results}
\label{sec:3_methods}

\subsection{General Gauge Theory}
\label{subsec:general-gauge-theory}
The one-loop results for the anomalous dimensions
$\gamma_{ab}(\mathrm{S})$, $\hat{\gamma}_{ab}(\mathrm{S})$ and $\beta$-functions
$\beta(v_a)$ in a general gauge theory have been presented
in~\cite{Sperling:2013eva} and read
\begin{subequations}
  \begin{align}
  \gamma_{a b}^{(1)}(\mathrm{S}) &= \frac{1}{(4 \pi)^2} \left[ g^2
    \left(3-\xi \right) \CasiScaGauge{a}{b} - \YukaInvGauge{a}{b} \right] \; ,
\\
   \hat{\gamma}_{a b}^{(1)}(\mathrm{S}) &= \frac{1}{(4 \pi)^2} 2 g^2 \xi \xi'
\CasiScaGauge{a}{b} \; ,\\
   \beta^{(1)}(v_{a}) &= \frac{1}{(4 \pi)^2} \left[ g^2 \left(3-\xi + 2
 \xi \xi'\right) \CasiScaGauge{a}{b} - \YukaInvGauge{a}{b} \right] v_{b} \;.
 \label{eqn:vev_beta_one-loop_generic}
  \end{align}
    \label{eqn:general_one-loop_results}%
\end{subequations}
At the two-loop level, the terms of $\mathcal{O}(g^2 YY^\dagger)$ of
$\hat{\gamma}^{(2)}$ \cite{Sperling:2013eva} and the full
$\gamma^{(2)}$ \cite{Machacek:1983tz,Luo:2002ti} have already been
published. Therefore, the computation of $\mathcal{O}(g^4)$-terms in
$\hat{\gamma}^{(2)}$ remains at two-loop.
Fig.~\ref{fig:TwoLoopGraphsZhat} contains the four relevant graphs
that generate the divergencies in the loop corrections of
$\Gamma_{\BrstBackGauge{a},K_{\ScaGauge{b}}}$, wherein we implicitly
understand one-loop subdivergencies to be subtracted. As before, all
calculations are carried out in \MSbar\ or equivalently MS scheme.
\begin{figure}[h]
  \centering
  \subfigure[]{%
    $\fmfvcenter{TwoPtBrstBackTwoLoopScaSelf1Vec1Sca}$
    \label{fig:TwoLoop1Vec1ScaInsertion}
  }%
  \qquad
  \subfigure[]{%
    $\fmfvcenter{TwoPtBrstBackTwoLoopGhoSelf1Vec1Gho}$
    \label{fig:TwoLoop1Vec1GhoInsertion}
  }%
  \qquad
  \subfigure[]{%
    $\fmfvcenter{TwoPtBrstBackTwoLoopVecExchange}$
    \label{fig:TwoLoop1VecExchange}
  }%
  \qquad
  \subfigure[]{%
    $\fmfvcenter{TwoPtBrstBackTwoLoopScaSelf2Spinor}$
    \label{fig:TwoLoop2SpinorInsertion}
  }%
  \caption{All relevant graphs for determination of two-loop corrections to
$\Gamma_{\BrstBackGauge{a},K_{\ScaGauge{b}}}$:
graphs~\ref{fig:TwoLoop1Vec1ScaInsertion}, \ref{fig:TwoLoop1Vec1GhoInsertion},
and \ref{fig:TwoLoop1VecExchange} are $\mathcal{O}(g^4)$-contributions;
graph~\ref{fig:TwoLoop2SpinorInsertion} corresponds to  $\mathcal{O}(g^2
YY^\dagger)$.}
  \label{fig:TwoLoopGraphsZhat}
\end{figure}

In analogy to the presentation of Machacek \&
Vaughn~\cite{Machacek:1983tz,Machacek:1983fi,Machacek:1984zw}, we
provide the contributions of each diagram of Fig.~\ref{fig:TwoLoopGraphsZhat}
in Tab.~\ref{tab:TwoLoop-Details} with the notation
\begin{align}
 \delta \hat{Z}_{ab}^{(2)} = \frac{1}{(4 \pi)^4} \hat{S}_{ab}\left(
\frac{A}{\eta^2} +\frac{B}{\eta} \right) \;,
\end{align}
wherein $1/\eta = 1/\epsilon + \ln(4 \pi) - \gamma_E$.
\begin{table}[h]
 \renewcommand\arraystretch{1.2}
  \begin{center}
  \begin{tabular}{cccc}
\toprule
$\qquad$ Diagram $\qquad$ & $\qquad \hat{S}_{ab} \qquad$ & $\qquad A \qquad$ &
$\qquad B \qquad$ \\ \hline
\ref{fig:TwoLoop1Vec1ScaInsertion} & $g^4 \xi \xi' \CasiScaGauge{a}{c}
\CasiScaGauge{c}{b}$ & $-3+\xi $ & $1+\xi $ \\
\ref{fig:TwoLoop1Vec1GhoInsertion} & $g^4 \xi \xi' \CasiAdjGauge{}{}
\CasiScaGauge{a}{b} $ & $\frac{-3 + \xi}{4} $ & $ \frac{1 + \xi}{4}  $ \\
\ref{fig:TwoLoop1VecExchange} & $g^4 \xi \xi' \CasiAdjGauge{}{}
\CasiScaGauge{a}{b} $ & $-\frac{\xi}{2} $ & $ \frac{3-\xi}{2} $ \\
\ref{fig:TwoLoop2SpinorInsertion} & $g^2 \xi \xi' \CasiScaGauge{a}{c}
\YukaInvGauge{c}{b} $ & $1 $ & $-1 $ \\
\bottomrule
\end{tabular}
\end{center}
\caption{Singular parts of the two-loop diagrams for
$\Gamma_{\BrstBackGauge{a},K_{\ScaGauge{b}}}$. All relevant one-loop
subdiagrams have been renormalized such that the above expressions
correspond to the two-loop diagrams depicted plus the necessary diagrams with
one-loop counterterm insertions.}
\label{tab:TwoLoop-Details}
\end{table}

The completed two-loop results in the \MSbar\ scheme read as follows
\begin{subequations}
\begin{align}
 \gamma_{a b}^{(2)} (\mathrm{S}) &= \frac{1}{(4 \pi )^4} \Bigg\{ g^4
\CasiScaGauge{a}{b}\left[\left(\frac{35}{3} - 2 \xi -\frac{1}{4} \xi^2 \right)
\CasiAdjGauge{}{} - \frac{10}{6} \DykinFermiGauge{}{} - \frac{11}{12}
\DykinScaGauge{}{} \right] %
  \label{eqn:gamma_2loop} \\*
 &\phantom{= \frac{1}{(4 \pi )^4} \Bigg\{}
 -\frac{3}{2}g^4\CasiScaGauge{a}{c}\CasiScaGauge{c}{b} +\frac{3}{2}
H_{ab}^{2}(\mathrm{S}) + \bar{H}_{ab}^{2}(\mathrm{S}) - \frac{10}{2} g^2
Y_{ab}^{2 F}(\mathrm{S}) - \frac{1}{2} \Lambda_{ab}^{2}(\mathrm{S}) \Bigg\} ,
\nonumber \\%
 \hat{\gamma}_{ab}^{(2)}(\mathrm{S}) &=
\frac{\xi \xi'}{(4 \pi)^4}\Bigg\{ g^4 \left[ 2 \left(1 + \xi \right)
\CasiScaGauge{a}{c} \CasiScaGauge{c}{b} +\frac{7-\xi}{2} \CasiAdjGauge{}{}
\CasiScaGauge{a}{b} \right]  %
  \label{eqn:gamma-hat_2loop} \\*%
 &\phantom{= \frac{\xi \xi'}{(4 \pi)^4}\Bigg\{ g^4 \bigg[ 2 \left(1 + \xi
\right) \CasiScaGauge{a}{c} \CasiScaGauge{c}{b} \;}
 - 2 g^2 \CasiScaGauge{a}{c} \YukaInvGauge{c}{b}  \Bigg\} \;, %
 \nonumber\\
 \BetaFkt{2}{v_a} &= \frac{1}{(4 \pi )^4} \Bigg\{ g^4
 \CasiScaGauge{a}{b}\left[\left(\frac{35}{3} - 2 \xi -\frac{1}{4} \xi^2 +
 \frac{7-\xi}{2}\xi \xi' \right)
 \CasiAdjGauge{}{} - \frac{10}{6} \DykinFermiGauge{}{} - \frac{11}{12}
 \DykinScaGauge{}{} \right] \nonumber %
 \label{eqn:vev_beta_two-loop_generic} \\*
  &\phantom{= \frac{1}{(4 \pi )^4} \Bigg\{}
  +g^4 \left[ 2 \xi \xi' \left(1 + \xi \right) -\frac{3}{2} \right]
 \CasiScaGauge{a}{c}\CasiScaGauge{c}{b}  - \frac{1}{2}
 \Lambda_{ab}^{2}(\mathrm{S})  \\*
 &\phantom{= \frac{1}{(4 \pi )^4} \Bigg\{} +\frac{3}{2} H_{ab}^{2}(\mathrm{S}) +
 \bar{H}_{ab}^{2}(\mathrm{S}) - \frac{10}{2} g^2 Y_{ab}^{2 F}(\mathrm{S})  - 2
 \xi \xi' g^2 \CasiScaGauge{a}{c} \YukaInvGauge{c}{b} \Bigg\} v_b \nonumber \; .
 \end{align}
\end{subequations}
\subsection{Kinetic Mixing}
\label{subsec:kin_mixing}
The results of Sec.~\ref{subsec:general-gauge-theory} hold for simple
gauge groups.  The generalization to product groups is obvious, except
for gauge kinetic mixing of $U(1)$ field strength tensors.  In the
recent literature, the impact of gauge kinetic mixing on RGEs has been
studied quite extensively up to two-loop
level~\cite{Luo:2002iq,Fonseca:2011vn,Fonseca:2013bua}.  Following the
approach of Refs.~\cite{Fonseca:2011vn,Fonseca:2013bua}, we need to
provide substitution rules for $\hat{\gamma}$ to take kinetic mixing
into account.

A generic gauge group $G$ can be decomposed into
  \begin{align}
   G = \bigg( \bigotimes_{k \in I} G_k \bigg)  \otimes \bigg( \bigotimes_{a \in
J} U(1)_a \bigg) \;,
  \end{align}
with the simple groups $G_k$ and the two (finite) sets $I,J \subset
\mathbb{N}$. 
The part of the Lagrangian describing kinetic mixing reads
\begin{align}
 \Lagr = -\frac{1}{4} \sum_{k \in I} F_{k,\mu \nu}^{A_k} F_{k}^{A_k,\mu
\nu} - \frac{1}{4} \sum_{a,b \in J} F_{a,\mu \nu} \Xi_{ab} F_{b}^{\mu \nu} +
\cdots \; .
\end{align}
Analogously to Refs.~\cite{Fonseca:2011vn,Fonseca:2013bua}, we define
\begin{align}
 \hat{g}_{ab}:= \sum_{c \in J} \delta_{ac} g'_c \sqrt{\Xi}_{cb}^{-1} \quad
\text{and} \quad W_a := \sum_{b \in J} Q_b \hat{g}_{ba} \; ,%
  \label{eqn:generalized_coupling_and_charge}
\end{align}
with the root defined by $\sqrt{\Xi} \sqrt{\Xi} = \Xi$.

The inspection of the graphs in Fig.~\ref{fig:TwoLoopGraphsZhat} implies
that there do not exist any gauge kinetic mixing contributions to
$\hat{\gamma}^{(1)}$ and the $\mathcal{O}(g^2 YY^\dagger)$-part of
$\hat{\gamma}^{(2)}$, because BRS-ghost and -antighost are not affected by
kinetic mixing. Graphs~\ref{fig:TwoLoop1Vec1GhoInsertion} and
\ref{fig:TwoLoop1VecExchange} are not affected either, as $U(1)$-gauge fields
do not interact with the corresponding Faddeev--Popov-ghosts. Hence, the only
change for kinetic mixing stems from graph~\ref{fig:TwoLoop1Vec1ScaInsertion},
in particular from the one-loop insertion of the scalar self-energy. The
relevant substitution rule is given by
\begin{align}
  g^4 \CasiScaGauge{}{} \CasiScaGauge{}{}
&\xrightarrow[\text{kin. mix}]{\hat{\gamma}} \left[
\sum_{k\in I} g_k^2 \CasiGenRep{G_k}{X} + \sum_{d\in J} W_d(X) W_d(X) \right] %
\label{eqn:Ersetzungregel_Zhat_kinMixing} \\
&\phantom{\xrightarrow[\text{kin. mix}]{\hat{\gamma}}
\bigg[ \sum_{k\in I} g_k^2 \CasiGenRep{G_k}{X} } \times \left[\sum_{k\in I}
g_k^2 \CasiGenRep{G_k}{X} + \sum_{d\in J} {g'}_d^2 Q_d^2(X)\right] \nonumber
\; .%
\end{align} 
Here $g_k$ denote the non-abelian gauge couplings and $g'_d$ the
abelian ones, with the corresponding quantum numbers $Q_d$. Further,
$X$ denotes the field under consideration, e.g. up- or down-type
Higgs.  The substitution rules for $\gamma$ can be found in
\cite{Fonseca:2011vn,Fonseca:2013bua}.
\subsection{Supersymmetric Gauge Theory}
The treatment of supersymmetric theories requires to take three
subtleties into account: (i) supersymmetric theories are formulated in
terms of complex scalar fields, (ii) the coupling structure is
severely restricted by supersymmetry, and (iii) the use of the
supersymmetry-preserving renormalization scheme \DRbar.

The first two points are merely computational issues, in the sense that one
needs to take care of the changed coupling structure and the scalar
field representation. Hence, these aspects will not be spelled out in detail and
we directly present the results for complex scalar fields in a notation based
on Ref.~\cite{Martin:2001vx}. We will, however, give some details on
the conversion to \DRbar, which requires
transition counterterms for parameters~\cite{Martin:1993yx} and
fields~\cite{Stockinger:2011gp}. The existence of such transition counterterms
is due to the equivalence of dimensional reduction and dimensional
regularisation as shown in Ref.\ \cite{Jack:1994bn}.

At one-loop level the results have been provided
earlier~\cite{Sperling:2013eva} and read 
\begin{subequations}
  \begin{align}
  \gamma_{ab}^{(1)}(\mathrm{S}) \Big|_{\mathrm{SUSY}}^{\text{\DRbar}} &=
\frac{1}{(4 \pi)^2} \left[ g^2 \left(1-\xi \right) \CasiScaGauge{a}{b} -
\frac{1}{2} Y_{apq}^{*} Y_{bpq}^{\phantom{*}} \right]   \; , %
  \label{eqn:gamma_1-Loop_SUSY-complex}\\ 
  \hat{\gamma}_{ab}^{(1)}(\mathrm{S}) \Big|_{\mathrm{SUSY}}^{\text{\DRbar}}
&= \frac{1}{(4 \pi)^2} 2 g^2 \xi \xi' \CasiScaGauge{a}{b}   \; ,%
\label{eqn:hat-gamma_1-Loop-SUSY-complex} \\
   \beta^{(1)}(v_{a}) \Big|_{\mathrm{SUSY}}^{\text{\DRbar}} &= \frac{1}{(4
 \pi)^2} \left[ g^2 \left(1-\xi + 2 \xi \xi'\right) \CasiScaGauge{a}{b} - 
 \frac{1}{2} Y_{apq}^{*} Y_{bpq}^{\phantom{*}} \right] v_{b} \; . %
   \label{eqn:beta_VEV_SUSY_1-loop-complex}%
  \end{align}%
  \label{eqn:SUSY_1-Loop_results-complex}%
\end{subequations}%
The first two-loop renormalization studies of softly broken $N=1$ SUSY theories
in \DRbar\ have been performed in
\cite{Martin:1993zk,Jack:1994kd,Yamada:1994id}, though not always in component
fields as used here. To our knowledge, the full result for $\gamma^{(2)}$ in a
general supersymmetric theory is not available in the literature, except for
Landau gauge ($\xi=0$) \cite{Martin:2001vx}.  In order to obtain the
result  for
arbitrary $\xi$ we proceed in the following steps.
We first reevaluate the Feynman graphs in Ref.~\cite{Machacek:1983tz} with
a generic $N=1$ supersymmetric Lagrangian.\footnote{Note the remarks by Ref.~\cite{Luo:2002ti} on
the implicitly real spinors of Machacek \& Vaughn.}
Then we apply transition counterterms for the conversion from \MSbar\
to \DRbar. This step differs from the case of the \DRbar\
$\beta$-functions computed in Ref.\ \cite{Martin:1993zk}. Since the
$\beta$-functions in that reference are gauge invariant, physical
quantities, 
only transition counterterms for physical parameters were required, and
those were provided in Ref.\ \cite{Martin:1993yx}. In the present case
of $\gamma$-functions, also transition counterterms for field
renormalization and gauge parameters are necessary. These were
presented in Ref.\ \cite{Stockinger:2011gp}. Fortunately, however, the
needed additional transition counterterms for the scalar field
renormalization and for the gauge parameter are zero,
\begin{align}
    \delta Z_{\varphi}^{(1),\text{trans}}=0, %
    \label{eqn:scalar_transition-CT}\\
 \delta Z_{\xi}^{(1),\text{trans}}=0.
  \label{eqn:xi_transition-CT}
\end{align}

The transition for $\hat{\gamma}$ to supersymmetry and \DRbar\
could be carried out in an analogous way, by employing transition
counterterms. However, it is also possible and simpler to use the fact
that there is no difference between \MSbar\ and \DRbar\ for any
diagram contributing to $\delta\hat{Z}$ at the two-loop level. Hence,
$\hat\gamma$ is equal in the \MSbar\ and \DRbar\ schemes. From this
knowledge, one can then derive additional transition counterterms as a
by-product: $\delta
\hat{Z}^{(1),\text{trans}}=0 $, and owing to the non-renormalization
of the gauge fixing,
\begin{align}
 \delta Z_{\xi'}^{(1),\text{trans}} = - \delta Z_{g}^{(1),\text{trans}}  +
\frac{1}{2} \delta Z_{V}^{(1),\text{trans}} = \frac{1}{(4 \pi)^2}
\frac{g^2}{3} \CasiAdjGauge{}{} \; , %
\label{eqn:xiPrim_transition-CT}
\end{align}
where $\delta Z_{V}^{(1),\text{trans}}$ denotes the transition
counterterm for the gauge 
field, as obtained in Ref.~\cite{Stockinger:2011gp}.
With these ingredients, the full
gauge-dependent two-loop results for the 
anomalous dimensions $\gamma$ and $\hat{\gamma}$ as well as for the VEV
$\beta$-function can be obtained. In \DRbar\ they read
\begin{subequations}
 \begin{align}
\gamma_{ab}^{(2)}(\mathrm{S})\Big|_{\mathrm{SUSY}}^{\text{\DRbar}} &=
\frac{1}{(4 \pi)^4} \bigg\{
   g^4 \left[ \left( \frac{9}{4} -\frac{5}{3} \xi -\frac{1}{4}  \xi^2
\right) \CasiAdjGauge{}{}  - \DykinScaGauge{}{}
\right] \CasiScaGauge{a}{b} %
\label{eqn:gamma_2-loop_DRbar-complex} \\*
&\phantom{= \frac{1}{(4 \pi)^4} \bigg\{ } -2 g^4 \CasiScaGauge{a}{c}
\CasiScaGauge{c}{b}
+ \frac{1}{2}  Y_{arc}^{*} Y_{rpq}^{\phantom{*}} Y_{pqd}^{*}
Y_{bcd}^{\phantom{*}}
  \nonumber \\*
&\phantom{= \frac{1}{(4 \pi)^4} \bigg\{ }   
+ g^2 \left[ \CasiScaGauge{a}{c} Y_{cpq}^{*} Y_{bpq}^{\phantom{*}} -2
  Y_{apq}^{*} \CasiScaGauge{p}{r} Y_{brq}^{\phantom{*}} \right]
 \bigg\}
\nonumber \; ,\\
\hat{\gamma}_{ab}^{(2)}(\mathrm{S})
\Big|_{\mathrm{SUSY}}^{\text{\DRbar/\MSbar}} &=
\frac{\xi \xi'}{(4 \pi)^4} \Bigg\{ g^4   \left[ \frac{7-\xi}{2}
\CasiAdjGauge{}{} \CasiScaGauge{a}{b} -2\left(1 - \xi \right)
\CasiScaGauge{a}{c} \CasiScaGauge{c}{b} \right] %
\label{eqn:hat-gamma_2-Loop-SUSY-complex} \\*
 &\phantom{=\frac{\xi \xi'}{(4 \pi)^4} \Bigg\{ g^4   \Bigg[-2\left(1 - \xi
\right) \CasiScaGauge{a}{a} }- g^2 \CasiScaGauge{a}{c} Y_{cpq}^{*}
Y_{bpq}^{\phantom{*}}  \Bigg\}  \nonumber  \; , \\%
  \beta^{(2)}(v_{a}) \Big|_{\mathrm{SUSY}}^{\text{\DRbar}} &=
 \frac{1}{(4 \pi)^4} \Bigg\{   g^4  \left[ \left(
 \frac{9}{4} -\frac{5}{3} \xi -\frac{1}{4}  \xi^2 + \frac{7-\xi}{2}\xi \xi'
 \right) \CasiAdjGauge{}{} - \DykinScaGauge{}{} \right] \CasiScaGauge{a}{b} %
  \\*
 &\phantom{=\frac{\xi \xi'}{(4 \pi)^4} \Bigg\{  } - g^4 \left[2 \xi \xi'\left(1
 - \xi \right)+ 2\right] \CasiScaGauge{a}{c} \CasiScaGauge{c}{b}  + \frac{1}{2} 
 Y_{arc}^{*} Y_{rpq}^{\phantom{*}} Y_{pqd}^{*}
 Y_{bcd}^{\phantom{*}} \nonumber \\*
 &\phantom{=\frac{\xi \xi'}{(4 \pi)^4} \Bigg\{  }
 + g^2 \left[ 1- \xi \xi'\right] \CasiScaGauge{a}{c} Y_{cpq}^{*}
 Y_{bpq}^{\phantom{*}} 
 - 2 g^2  Y_{apq}^{*} \CasiScaGauge{p}{r}
 Y_{brq}^{\phantom{*}}  \Bigg\} \; v_{b} \; . \nonumber %
 \end{align}%
\label{eqn:SUSY_2-Loop_results-complex}%
\end{subequations}%

\section{Application to Concrete Supersymmetric Models}
\label{sec:4_results}
This section provides the explicit two-loop results for the
renormalization of all VEVs in the MSSM,
NMSSM, and \ESSM, using the notation of Ref.\
\cite{Sperling:2013eva}. For completeness and convenience, we provide
the full results including previously known ones.
\subsection{MSSM}
\label{subsec:MSSM}
\paragraph{one-loop}
The one-loop results for the anomalous dimensions of the MSSM Higgs
doublets read 
\begin{subequations}
 \begin{align}
 (4 \pi)^2 \gamma_{\text{MSSM}}^{(1),\text{\DRbar}}(H_{u})&=  \left(1-\xi 
\right)
\left( \frac{3}{20} g_1^2 + \frac{3}{4} g_2^2 \right) - N_c \Tr \left( y^{u}
y^{u \dagger} \right)  \; , \\
 (4 \pi)^2 \hat{\gamma}_{\text{MSSM}}^{(1),\text{\DRbar}}(H_{u})&=  2 \xi
\xi' \left( \frac{3}{20} g_1^2 + \frac{3}{4} g_2^2 \right)\; .
  \end{align}%
  \label{eqn:anoDim_Hu_MSSM_1-Loop}%
\end{subequations}%
\begin{subequations}
  \begin{align}
 (4 \pi)^2 \gamma_{\text{MSSM}}^{(1),\text{\DRbar}}(H_{d})&= \left(1-\xi
\right) \left( \frac{3}{20} g_1^2 + \frac{3}{4} g_2^2 \right) - N_c\Tr \left(
y^{d} y^{d \dagger} \right) - \Tr \left( y^{e} y^{e \dagger} \right)  \; , \\
 (4 \pi)^2 \hat{\gamma}_{\text{MSSM}}^{(1),\text{\DRbar}}(H_{d})&=  2
\xi \xi' \left( \frac{3}{20} g_1^2 + \frac{3}{4} g_2^2 \right) \; .
  \end{align}%
 \label{eqn:anoDim_Hd_MSSM_1-Loop}%
\end{subequations}%
The $\beta$-function of $\tan \beta$ follows then as
\begin{align}
  \frac{\beta_{\text{MSSM}}^{(1),\text{\DRbar}}(\tan\beta)}{\tan\beta} = -
\frac{1}{(4 \pi)^2}    \left[ N_c \Tr \left( y^{u} y^{u \dagger} \right)  - N_c
\Tr \left( y^{d} y^{d \dagger} \right) -  \Tr \left( y^{e} y^{e \dagger}
\right)  \right] \;. %
  \label{eqn:tanbeta_MSSM_1-Loop_final}%
\end{align}%
\paragraph{two-loop}
The application of the general two-loop results yields for the MSSM
\begin{subequations}%
 \begin{align}%
 (4 \pi)^4
\gamma_{\text{MSSM}}^{(2),\text{\DRbar}}(H_{u})&=-\frac{207}{200}g_1^4
-\frac{9}{20} g_1^2 g_2^2 - \left( 3 +\frac{5}{2} \xi  + \frac{3}{8}
\xi^2  \right)g_2^4 %
\label{eqn:anoDim_Hu_MSSM_2Loop}\\*
&\phantom{=\;}- \left(\frac{4}{15} g_1^2 + \frac{16}{3} g_3^2 \right) N_c \Tr
\left(y^u y^{u \dagger}\right) \nonumber\\*
&\phantom{=\;}+N_c \Tr \left(y^u y^{d\dagger} y^d y^{u \dagger}\right)+ 3 N_c
\Tr \left(y^u y^{u\dagger} y^u y^{u\dagger}\right) \nonumber \; , \\
 (4 \pi)^4 \hat{\gamma}_{\text{MSSM}}^{(2),\text{\DRbar}}(H_{u}) &=-  \xi \xi'
\left\{ \left( \frac{3}{10} g_1^2 + \frac{3}{2} g_2^2 \right) \left[ N_c \Tr
\left( y^{u} y^{u \dagger} \right) \right]  + \mathcal{R}_{\text{MSSM}} \right\}
  \label{eqn:anoDimHat_Hu_MSSM_2Loop} \;,
 \end{align}%
  \label{eqn:anoDim_Hu_MSSM_2-Loop}%
\end{subequations}%
\begin{subequations}%
  \begin{align}%
  (4 \pi)^4 \gamma_{\text{MSSM}}^{(2),\text{\DRbar}}(H_{d})&=
-\frac{207}{200}g_1^4
-\frac{9}{20} g_1^2 g_2^2 - \left( 3 +\frac{5}{2} \xi  + \frac{3}{8}
\xi^2  \right)g_2^4%
\label{eqn:anoDim_Hd_MSSM_2Loop}\\*
&\phantom{=\;}- \left( -\frac{2}{15} g_1^2 + \frac{16}{3} g_3^2 \right)
N_c\Tr\left(y^d y^{d\dagger} \right)-\frac{6}{5} g_1^2 \Tr \left(y^e
y^{e\dagger}\right) \nonumber \\*
&\phantom{=\;}+3 N_c \Tr \left(y^d y^{d\dagger}y^d y^{d\dagger}\right)+N_c
\Tr \left(y^d y^{u\dagger} y^u y^{d\dagger}\right)+3
\Tr \left(y^e y^{e\dagger}y^e y^{e\dagger}\right) \nonumber \; , \\
 (4 \pi)^4 \hat{\gamma}_{\text{MSSM}}^{(2),\text{\DRbar}}(H_{d}) &=- \xi \xi'
\left\{
\left( \frac{3}{10} g_1^2 + \frac{3}{2} g_2^2 \right) \left[N_c \Tr \left( y^{d}
y^{d \dagger} \right) +\Tr \left( y^{e} y^{e \dagger} \right) \right]
+\mathcal{R}_{\text{MSSM}}  \right\} %
  \label{eqn:anoDimHat_Hd_MSSM_2Loop} \;,%
  \end{align}%
  \label{eqn:anoDim_Hd_MSSM_2-Loop}%
\end{subequations}%
with
\begin{align}
  \mathcal{R}_{\text{MSSM}}&= (1-\xi) \frac{9}{2}\left( \frac{1}{100} g_1^4 +
    \frac{1}{10} g_1^2 g_2^2 + \frac{1}{4} g_2^4  \right) - 3\frac{7-\xi}{4}
  g_2^4 \; . %
 \label{eqn:RTerm_MSSM}
\end{align}
The explicit calculations confirm our earlier statement~\cite{Sperling:2013eva}
that the same
$\mathcal{R}_{\text{MSSM}}$ terms in $\hat{\gamma}^{(2)}$ appear for
up- and down-Higgs. Thus, we obtain the two-loop $\beta$-function for
$\tan \beta$ as
\begin{align}%
 \frac{\beta_{\text{MSSM}}^{(2),\text{\DRbar}}(\tan\beta)}{\tan\beta} &= 
\frac{1}{(4
\pi)^4} \Bigg\{  - \left(\frac{4}{15} g_1^2 + \frac{16}{3} g_3^2 \right) N_c
\Tr \left(y^u y^{u \dagger}\right) %
\label{eqn:beta_tan_two-loop_MSSM} \\*
&\phantom{=  \frac{1}{(4 \pi)^4} \Bigg\{ } + \left( -\frac{2}{15} g_1^2 +
\frac{16}{3} g_3^2 \right) N_c\Tr\left(y^d y^{d\dagger} \right) +\frac{6}{5}
g_1^2 \Tr \left(y^e y^{e\dagger}\right) \nonumber\\*
&\phantom{=  \frac{1}{(4 \pi)^4} \Bigg\{ } +  3 N_c \Tr \left(y^u y^{u\dagger}
y^u y^{u\dagger}\right) -3 N_c \Tr \left(y^d y^{d\dagger}y^d
y^{d\dagger}\right)- 3
\Tr \left(y^e y^{e\dagger}y^e y^{e\dagger}\right) \Bigg\} \nonumber\\
&\phantom{=\;}+ \frac{1}{(4 \pi)^2} \xi \xi' \left( \frac{3}{10}
g_1^2 + \frac{3}{2} g_2^2 \right) 
\frac{\beta_{\text{MSSM}}^{(1),\text{\DRbar}}(\tan\beta)}{\tan\beta}  \;
,\nonumber%
\end{align}%
The gauge-dependence of $\tan \beta$ at two-loop stems solely from the
$\hat{\gamma}$ terms. 
\subsection{NMSSM}
\label{subsec:NMSSM}
\paragraph{one-loop}
The one-loop anomalous dimensions for the Higgs doublets $H_{u,d}$ in
the NMSSM resemble the corresponding MSSM results: 
\begin{subequations}%
  \begin{align}%
   \gamma_{\text{NMSSM}}^{(1),\text{\DRbar}}(H_{u,d}) &=
\gamma_{\text{MSSM}}^{(1),\text{\DRbar}}(H_{u,d}) -
\frac{1}{(4 \pi)^2} |\lambda|^2 \; , \\
  \hat{\gamma}_{\text{NMSSM}}^{(1),\text{\DRbar}}(H_{u,d}) &=
 \hat{\gamma}_{\text{MSSM}}^{(1),\text{\DRbar}}(H_{u,d}) \; .
  \end{align}%
    \label{eqn:anoDim_Hu-Hd_NMSSM_1-Loop}%
\end{subequations}%
The NMSSM Higgs singlet $S$ has the following RGE coefficients: 
\begin{subequations}%
  \begin{align}%
   \gamma_{\text{NMSSM}}^{(1),\text{\DRbar}}(S) &=  - \frac{1}{(4 \pi)^2} 2
\left(|\lambda|^2 + |\kappa|^2 \right) \; , \\
   \hat{\gamma}_{\text{NMSSM}}^{(1),\text{\DRbar}}(S) &=0 \; .
  \end{align}%
  \label{eqn:anoDim_S_NMSSM_1-Loop}%
\end{subequations}%
Due to the unchanged gauge group the one-loop result for $\tan \beta$
is identical to the MSSM
\begin{align}%
 \beta_{\text{NMSSM}}^{(1),\text{\DRbar}}(\tan \beta) =
\beta_{\text{MSSM}}^{(1),\text{\DRbar}}(\tan \beta) \; .%
  \label{eqn:tanbeta_NMSSM_1-Loop_final}%
\end{align}%
\paragraph{two-loop}
The two-loop results for the Higgs-doublets are given by
\begin{subequations}
 \begin{align}
 \gamma_{\text{NMSSM}}^{(2),\text{\DRbar}}(H_{u})&=
\gamma_{\text{MSSM}}^{(2),\text{\DRbar}}(H_{u}) +
\frac{|\lambda|^2}{ (4 \pi)^4} \left[ 2 |\kappa|^2  + 3 |\lambda|^2 +  N_c \Tr
\left(y^d y^{d\dagger}\right)+  \Tr \left(y^e y^{e\dagger}\right) \right] \;,\\
   \hat{\gamma}_{\text{NMSSM}}^{(2),\text{\DRbar}}(H_{u}) &=  -\frac{\xi
\xi'}{(4 \pi)^4} \Bigg\{ \left( \frac{3}{10} g_1^2 + \frac{3}{2} g_2^2
\right) \left[ N_c \Tr \left( y^{u} y^{u \dagger} \right) + |\lambda|^2 \right]
+ \mathcal{R}_{\text{NMSSM}} \Bigg\} \; ,
 \end{align}%
  \label{eqn:anoDim_Hu_NMSSM_2-Loop}%
\end{subequations}%
\begin{subequations}
 \begin{align}
   \gamma_{\text{NMSSM}}^{(2),\text{\DRbar}}(H_{d})&=
\gamma_{\text{MSSM}}^{(2),\text{\DRbar}}(H_{d})+
\frac{|\lambda|^2}{ (4 \pi)^4} \left[ 2 |\kappa|^2  + 3 |\lambda|^2 +  N_c \Tr
\left(y^u y^{u\dagger}\right) \right] \;,\\
 \hat{\gamma}_{\text{NMSSM}}^{(2),\text{\DRbar}}(H_{d}) &= -\frac{\xi \xi'}{(4
\pi)^4} \Bigg\{ \left( \frac{3}{10} g_1^2 + \frac{3}{2} g_2^2
\right) \left[ N_c \Tr \left( y^{d} y^{d \dagger} \right) +\Tr \left( y^{e}
y^{e \dagger} \right) + |\lambda|^2 \right]  \\*
  &\phantom{=  -\frac{\xi \xi'}{(4 \pi)^4} 
\Bigg\{ \left( \frac{3}{10} g_1^2 + \frac{3}{2} g_2^2
\right) \left[ N_c \Tr \left( y^{u} y^{u \dagger} \right) + |\lambda|^2 \right]}
 + \mathcal{R}_{\text{NMSSM}} \Bigg\} \nonumber \; ,%
 \end{align}%
  \label{eqn:anoDim_Hd_NMSSM_2-Loop}%
\end{subequations}%
with $\mathcal{R}_{\text{NMSSM}} = \mathcal{R}_{\text{MSSM}}$.  Again,
the $\mathcal{R}_{\text{NMSSM}}$ terms in $\hat{\gamma}^{(2)}$ are
equal for up- and down-Higgs.  Next, we can provide the results for
the two-loop gauge singlet:
\begin{subequations}%
  \begin{align}%
 (4 \pi)^4 \gamma_{\text{NMSSM}}^{(2),\text{\DRbar}}(S)&=  8 |\kappa|^4 +8
|\kappa|^2 |\lambda|^2 + 4 |\lambda|^4 - \left(\frac{6}{5}  g_1^2 + 6 g_2^2
\right) |\lambda|^2 \\*
&\phantom{=\;}+ 2 |\lambda|^2 \left[ N_c \Tr
\left(y^d y^{d\dagger}\right)+ \Tr \left(y^e y^{e\dagger}\right)+
 N_c \Tr \left(y^u y^{u\dagger}\right) \right]  \nonumber \;, \\
 \hat{\gamma}_{\text{NMSSM}}^{(2),\text{\DRbar}}(S) &=0 \; .%
  \end{align}%
    \label{eqn:anoDim_S_NMSSM_2-Loop}%
\end{subequations}%
Finally, the two-loop $\beta$-function for $\tan\beta$ turns out to be
modified by the additional Yukawa-coupling $\lambda$ in comparison to
the MSSM
\begin{subequations}
\begin{align}
 \frac{ \beta_{\text{NMSSM}}^{(2),\text{\DRbar}}(\tan\beta)}{\tan\beta} &=
\gamma_{\text{MSSM}}^{(2),\text{\DRbar}}(H_{u}) -
\gamma_{\text{MSSM}}^{(2),\text{\DRbar}}(H_{d}) +
\frac{|\lambda|^2}{ (4 \pi)^2} \frac{
\beta_{\text{MSSM}}^{(1),\text{\DRbar}}(\tan\beta)}{\tan
\beta} \\*
&\phantom{=\;}+ \frac{1}{(4 \pi)^2} \xi \xi' \left( \frac{3}{10}
g_1^2 + \frac{3}{2} g_2^2 \right) \frac{
\beta_{\text{MSSM}}^{(1),\text{\DRbar}}(\tan\beta)}{\tan \beta}   \nonumber
\\
&= \frac{ \beta_{\text{MSSM}}^{(2),\text{\DRbar}}(\tan\beta)}{\tan\beta}  +
\frac{|\lambda|^2}{ (4 \pi)^2} \frac{
\beta_{\text{MSSM}}^{(1),\text{\DRbar}}(\tan\beta)}{\tan \beta} \; .%
  \end{align}%
\label{eqn:beta_tan_two-loop_NMSSM}%
\end{subequations}%
\subsection{\ESSM}
The \ESSM\ introduces a new feature: The
$U(1)_N$-extension of the SM-gauge group leads inevitably to gauge
kinetic mixing.  The notations for kinetic mixing of
Sec.~\ref{subsec:kin_mixing} can be specialized to the \ESSM\ as
\begin{align}
 \hat{g} = \begin{pmatrix} g_1 & g_{11'} \\ g_{1'1} & g'_1 \end{pmatrix}
\quad \text{and} \quad  Q(X) := \begin{pmatrix}
\sqrt{\frac{3}{5}} Q_Y(X) \\ \sqrt{\frac{1}{40}} Q_N(X)  \end{pmatrix} \;. %
  \label{eqn:mixing-E6SSM}%
\end{align}%
Note that Eq.~\eqref{eqn:mixing-E6SSM} contains the GUT-normalized
$U(1)_Y$- and $U(1)_N$-charges for any field $X$. The quantum-numbers $Q_Y(X)$
and $Q_N(X)$ are those of Ref.~\cite{King:2005jy}.
\paragraph{one-loop}
In comparison to our earlier results~\cite{Sperling:2013eva} the one-loop
anomalous dimensions $\gamma$ and $\hat{\gamma}$ are now extended for
the general case of gauge kinetic
mixing already present at tree-level. For the
Higgs-doublets $H_{u/d,3}$ and the SM-singlet $S_3$ our computations yield
\begin{subequations}%
  \begin{align}%
 \gamma_{\text{\ESSM}}^{(1),\text{\DRbar}}(H_{u,3})
&=\gamma_{\text{MSSM}}^{(1),\text{\DRbar}}(H_{u})
+\frac{1}{(4 \pi)^2} \left[ \frac{1}{10}(1-\xi) {g'_1}^2  -|\lambda_3|^2
\right] \\*
&\phantom{= \;}+\frac{1-\xi}{(4 \pi)^2} \left(\frac{3}{20} g_{11'}^2 +
\frac{1}{10} g_{1'1}^2 -\frac{1}{5} \sqrt{\frac{3}{2}} g_{11'} g'_1 
-\frac{1}{5} \sqrt{\frac{3}{2}} g_{1'1} g_1  \right) \nonumber \; , \\
\hat{\gamma}_{\text{\ESSM}}^{(1),\text{\DRbar}}(H_{u,3})
&=\hat{\gamma}_{\text{MSSM}}^{(1),\text{\DRbar}}(H_{u}) + \frac{1}{(4 \pi)^2}
\frac{1}{5} \xi \xi' {g'_1}^2  \; .
  \end{align}%
   \label{eqn:anoDim_Hu_ESSM_1-Loop}%
\end{subequations}%
\begin{subequations}%
  \begin{align}%
 \gamma_{\text{\ESSM}}^{(1),\text{\DRbar}}(H_{d,3})
&=\gamma_{\text{MSSM}}^{(1),\text{\DRbar}}(H_{d})
+\frac{1}{(4 \pi)^2} \left[ \frac{9}{40} \left(1-\xi \right) {g'_1}^2 -
|\lambda_3|^2 \right]  \\*
&\phantom{= \;}+\frac{1-\xi}{(4 \pi)^2} \left(\frac{3}{20} g_{11'}^2 +
\frac{9}{40} g_{1'1}^2 +\frac{3}{10} \sqrt{\frac{3}{2}} g_{11'} g'_1 
+\frac{3}{10} \sqrt{\frac{3}{2}} g_{1'1} g_1  \right) \nonumber \; , \\
\hat{\gamma}_{\text{\ESSM}}^{(1),\text{\DRbar}}(H_{d,3})
&=\hat{\gamma}_{\text{MSSM}}^{(1),\text{\DRbar}}(H_{d}) + \frac{1}{(4 \pi)^2} 
\frac{9}{20} \xi \xi' {g'_1}^2   \; .
  \end{align}%
  \label{eqn:anoDim_Hd_ESSM_1-Loop}%
\end{subequations}%
\begin{subequations}%
  \begin{align}%
 (4 \pi)^2 \gamma_{\text{\ESSM}}^{(1),\text{\DRbar}}(S_3) &= \frac{5}{8}
\left(1-\xi  \right) \left( {g'_1}^2 + g_{1'1}^2 \right)  - 2 \Tr \left(\lambda
\lambda^\dagger \right) - 
N_c \Tr \left( \kappa \kappa^\dagger \right)  \; , \\
(4 \pi)^2 \hat{\gamma}_{\text{\ESSM}}^{(1),\text{\DRbar}}(S_3) &= \frac{5}{4}
\xi \xi'{g'_1}^2 \; .
  \end{align}%
  \label{eqn:anoDim_S_ESSM_1-Loop}%
\end{subequations}%
Thus, the one-loop $\beta$-function for $\tan \beta$ is given by
  \begin{align}%
 \frac{\beta_{\text{\ESSM}}^{(1),\text{\DRbar}}(\tan\beta) }{\tan\beta}
 &= \frac{\beta_{\text{MSSM}}^{(1),\text{\DRbar}}(\tan\beta) }{\tan\beta} -
\frac{1}{(4 \pi)^2} \frac{1}{8} \left(1-\xi  + 2\xi \xi'\right) {g'_1}^{2} %
 \label{eqn:tan_beta_one-loop_ESSM} \\*
&\phantom{= \;} -\frac{1-\xi}{(4 \pi)^2} \left[ \frac{1}{8} g_{1'1}^2
+\frac{1}{2} \sqrt{\frac{3}{2}}  \left( g_{11'} g'_1 + g_{1'1} g_1 \right)
\right]  \; . \nonumber%
  \end{align}%
Eq.~\eqref{eqn:tan_beta_one-loop_ESSM} illustrates once more the
gauge dependence of $\tan \beta$ at one-loop level due to the different
$U(1)_N$-quantum numbers of the Higgs doublets, see
\cite{Sperling:2013eva}.
\paragraph{two-loop}
We restrict the list of two-loop results to the $\hat\gamma$ and the
$\beta$-function for $\tan\beta$. The two-loop results for the \ESSM\
Higgs doublets are 
\begin{subequations} 
  \begin{align}
   (4 \pi)^4 \hat{\gamma}_{\text{\ESSM}}^{(2),\text{\DRbar}}(H_{u,3}) &=  -\xi
\xi' \Bigg\{
\left( \frac{3}{10} g_1^2 + \frac{3}{2} g_2^2  + \frac{1}{5} {g'_1}^2\right)
\left[ N_c \Tr \left( y^{u} y^{u\dagger} \right) + |\lambda_3|^2 \right] +
\mathcal{R}_u \Bigg\} \; , \\
 (4 \pi)^4 \hat{\gamma}_{\text{\ESSM}}^{(2),\text{\DRbar}}(H_{d,3}) &=  - \xi
\xi' \Bigg\{ \left( \frac{3}{10} g_1^2 + \frac{3}{2} g_2^2 + \frac{9}{20}
{g'_1}^2 \right)  \\*
&\phantom{ - \xi \xi' \Bigg\{ \Bigg( \frac{3}{10} g_1^2 +
\frac{3}{2} g_2^2} \times \left[ N_c \Tr \left( y^{d} y^{d\dagger} \right) +\Tr
\left( y^{e} y^{e \dagger}  \right) + |\lambda_3|^2 \right] + \mathcal{R}_d
\Bigg\} \nonumber \; ,
 \end{align}%
  \label{eqn:anoDim_Hu_ESSM_2-Loop}%
\end{subequations}
with
\begin{subequations} 
  \begin{align}
\mathcal{R}_{u} &= \mathcal{R}_{\text{MSSM}} + (1-\xi)
\frac{1}{10} {g'_1}^2 \left[ \frac{3}{5} g_1^2  + 3 
g_2^2 + \frac{1}{5} {g'_1}^2 \right] \\*%
&\phantom{= \;}+ (1-\xi ) \frac{1}{200} \left[3 g_{11'}^2+2 g_{1'1}^2
-2 \sqrt{6} \left( g_{11'} g'_1 + g_{1'1} g_1\right) \right] \left(2 {g'_1}^2
+3 g_1^2+ 15 g_2^2 \right) %
\nonumber \; , \\
\mathcal{R}_{d} &= \mathcal{R}_{\text{MSSM}} + (1-\xi) \frac{9}{40}  {g'_1}^2
\left[ \frac{3}{5} g_1^2  + 3 g_2^2 + \frac{9}{20} {g'_1}^2 \right] \\
&\phantom{= \;}+ (1-\xi ) \frac{9}{800}  \left[2 g_{11'}^2 +3 g_{1'1}^2 +2
\sqrt{6} \left( g_{11'} g'_1 + g_{1'1} g_1 \right) \right] \left(3
{g'_1}^2 +2 g_1^2+ 10 g_2^2\right) \; . \nonumber%
 \end{align}%
  \label{eqn:anoDim_Hd_ESSM_2-Loop}%
\end{subequations}%
The new result of Eqs.~\eqref{eqn:anoDim_Hu_ESSM_2-Loop} and
\eqref{eqn:anoDim_Hd_ESSM_2-Loop} are the $\mathcal{R}$-terms for up-
and down-type Higgs. They differ non-trivially because of the
$U(1)_N$-quantum numbers, and $\mathcal{R}_u -\mathcal{R}_d$ does not
vanish in the \ESSM\ in contrast to the MSSM and NMSSM cases.  The
two-loop $\hat{\gamma}$ for the singlet field reads
\begin{subequations} 
  \begin{align}
(4 \pi)^4 \hat{\gamma}_{\text{\ESSM}}^{(2),\text{\DRbar}}(S_3) &=  -
\xi \xi' \left\{  \frac{5}{4} {g'_1}^2 \left[ 2 \Tr \left(\lambda
\lambda^\dagger \right) + N_c \Tr \left( \kappa \kappa^\dagger \right) \right] 
+ \mathcal{R}_s \right\} \; ,\\
\mathcal{R}_{s} &= \frac{25}{32} (1-\xi) {g'_1}^2 \left( {g'_1}^2 +
 g_{1'1}^2 \right)   \;.%
  \end{align}%
  \label{eqn:anoDim_S_ESSM_2-Loop}%
\end{subequations}%
The complete two-loop $\beta$-function of $\tan \beta$ requires additionally
the two-loop $\gamma$'s, which can be computed but will not be spelled out
here. The RGE coefficients then reads
\begin{align}
(4 \pi)^4 \frac{
\beta_{\text{\ESSM}}^{(2),\text{\DRbar}}(\tan\beta)}{\tan\beta} &=
(4 \pi)^4 \frac{ \beta_{\text{MSSM}}^{(2),\text{\DRbar}}(\tan\beta)}{\tan\beta}
+ (4 \pi)^2 |\lambda_3|^2  \frac{\beta^{(1),\text{\DRbar}}_{\text{MSSM}} (\tan
\beta)}{\tan\beta} %
 \label{eqn:beta_tan_two-loop_ESSM} \\*
&\phantom{= \;}
+ \frac{3}{40}\left[ -3 + \xi \xi' \left(1-\xi \right)   \right] {g'_1}^2 g_1^2
+ \frac{3}{8} \left[ 1+ \xi \xi' \left(1-\xi \right)  \right] {g'_1}^2 g_2^2 
\nonumber \\
&\phantom{= \;}
+\frac{1}{160} \left[ 201+13 \xi \xi' \left( 1-\xi \right) \right] {g'_1}^4
\nonumber \\
&\phantom{= \;}
-  \frac{1}{5} \left(1 -\frac{9}{4} \xi \xi' \right)  {g'_1}^2 
  \left[3 \Tr \left(y^d y^{d\dagger}\right) +\Tr \left(y^e y^{e\dagger}\right)
\right] \nonumber \\
&\phantom{= \;}
+ \frac{3}{10} \left(1-2 \xi \xi' \right) {g'_1}^2   \Tr \left(y^u
y^{u\dagger}\right) - \frac{1}{2} \left( 1- \frac{1}{2} \xi \xi' \right) 
{g'_1}^2 |\lambda_3|^2 \nonumber \\
&\phantom{= \;}
+\frac{3}{40} \left[ 11 +\frac{1}{2} \xi \xi' \left( 1 - \xi \right) \right]
\left( g_{11'}^2 {g'_1}^2 + g_{1'1}^2 g_1^2 \right) \nonumber \\
&\phantom{= \;}
+ \frac{1}{80} \left[ 201 +\frac{13}{2} \xi \xi' \left(1-\xi
\right) \right] g_{1'1}^2 {g'_1}^2  
+\frac{3}{8}\left[ 1 +\frac{1}{2} \xi' \xi \left(1-\xi \right)
\right] g_{1'1}^2 g_2^2 \nonumber \\
&\phantom{= \;}
+ \frac{1}{20} \sqrt{\frac{3}{2}} \left[ 99 +\frac{7}{2} \xi \xi' \left(1-\xi
\right) \right] \left( g_{11'} {g'_1} + g_{1'1}  g_1 \right) {g'_1}^2
\nonumber \\
&\phantom{= \;}
+ \frac{1}{10} \sqrt{\frac{3}{2}} \left[ 51 +\frac{3}{2} \xi
\xi' \left(1-\xi \right) \right] \left(  g_{11'} g'_1  +  g_{1'1}
g_1\right) g_1^2
\nonumber \\
&\phantom{= \;}
+\frac{3}{2} \sqrt{\frac{3}{2}}\left[ 1 +\frac{1}{2}  \xi \xi'
\left(1-\xi \right) \right] \left( g_{11'} g'_1 +g_{1'1} g_1   \right) g_2^2
\nonumber \\
&\phantom{= \;}
+\frac{51}{10} \sqrt{\frac{3}{2}} \left(g_{11'} g'_1 + g_{1'1} g_1 \right)
g_{11'}^2
 +\frac{99}{20} \sqrt{\frac{3}{2}} \left( g_{11'}  g'_1 
 +g_{1'1}  g_1 \right) g_{1'1}^2
 \nonumber \\
&\phantom{= \;} 
+\frac{21}{10} g_{11'} g_{1'1} g'_1 g_1
+\frac{201}{160} g_{1'1}^4 
-\frac{9}{40} g_{11'}^2 g_{1'1}^2 \nonumber\\
&\phantom{= \;}
- \left[ \frac{1}{2} g_{1'1}^2  
+\sqrt{6} \left( g_{11'} g'_1  + g_{1'1} g_1 \right) \right] |\lambda_3|^2 
\nonumber \\
&\phantom{= \;}
- \left[ \frac{2}{5} g_{11'}^2 
+\frac{3}{5} g_{1'1}^2 
+\frac{2}{5} \sqrt{6} \left( g_{11'} g'_1 + g_{1'1} g_1\right)  
 \right] \Tr \left(y^d y^{d\dagger}\right) \nonumber \\
&\phantom{= \;}
-\left[-\frac{6}{5} g_{11'}^2 
+\frac{1}{5} g_{1'1}^2  
+\frac{3}{5} \sqrt{\frac{3}{2}} \left( g_{11'} g'_1 + g_{1'1} g_1  \right) 
 \right] \Tr \left(y^e y^{e\dagger}\right) \nonumber \\
&\phantom{= \;}
-\left[ +\frac{4}{5} g_{11'}^2 
-\frac{3}{10} g_{1'1}^2 
+\frac{3}{5} \sqrt{\frac{3}{2}} \left( g_{11'} g'_1 + g_{1'1} g_1  \right) 
\right] \Tr \left(y^u y^{u\dagger}\right) \nonumber \; .%
\end{align}%
The connection with the more conventional
treatment~\cite{Babu:1996vt,King:2005jy} of the kinetic mixing in the \ESSM
\begin{subequations}
  \begin{align}
 \Lagr = -\frac{1}{4} F_Y^{\mu \nu} F_{Y,\mu \nu}^{\phantom{\mu}} 
-\frac{1}{4} F_N^{\mu \nu} F_{N,\mu \nu}^{\phantom{\mu}} -\frac{\sin{\chi}}{2}
F_Y^{\mu \nu} F_{N,\mu \nu}^{\phantom{\mu}} + \cdots %
  \label{eqn:Parametrisierung_Mixing}%
  \end{align}%
is established by the coupling matrix (c.f.
Eq.~\eqref{eqn:mixing-E6SSM})
  \begin{align}
\hat{g}= \begin{pmatrix}  g_1 & -g_1 \tan{\chi} \\  0 & \frac{g'_1}{\cos{\chi}} 
\end{pmatrix} \;.
  \end{align}
\end{subequations}

\section{Conclusions}
We completed the calculation of the two-loop VEV $\beta$-functions for
general gauge theories and generic supersymmetric theories.  The result
complements the well-known set of RGE coefficients of
Refs.~\cite{Machacek:1983tz,Machacek:1983fi,Machacek:1984zw,Luo:2002ti}
for general gauge theories as well as the supersymmetric gauge
theories of Refs.~\cite{Martin:1993zk,Martin:2001vx}.  In particular,
we achieved the following
\begin{itemize}
  \item Completion of $\hat{\gamma}^{(2)}$ by the missing
$\mathcal{O}(g^4)$-contributions of our earlier results~\cite{Sperling:2013eva}.
  \item Extension of $\gamma^{(2)}\big|_{\mathrm{SUSY}}^{\DRbar}$ to arbitrary
values of the gauge fixing parameter $\xi$.
\end{itemize}
As a consequence, we were able to provide the full VEV
$\beta$-function for general and supersymmetric gauge theories in the
\MSbar\ and \DRbar\ scheme up to the two-loop level.  The result was
applied to the MSSM, NMSSM, and \ESSM\ and we proved the statements
made in \cite{Sperling:2013eva} on the $\mathcal{O}(g^4)$-terms:
\begin{enumerate}
\item $\mathcal{R}_u -\mathcal{R}_d =0$ in the MSSM and NMSSM,
\item $\mathcal{R}_u -\mathcal{R}_d \neq 0$ for the \ESSM.
\end{enumerate}

\section*{Acknowledgments}
The authors thank Steve Martin and Florian Staub for discussions and
comments related to this work.  A.V.\ is supported by the German
Research Foundation DFG, STO/876/2-1.

\bibliographystyle{h-physrev}     
\footnotesize{\bibliography{Literatur}}      

\end{document}